 \documentclass[final,3p,times,twocolumn]{elsarticle}

\usepackage{graphics}
\usepackage{graphicx}

\usepackage{amssymb}
\usepackage[ansinew]{inputenc}
\usepackage{times} 
\usepackage{amsthm}
\usepackage{amsmath}
 \usepackage{lineno}

\journal{Nonlinear Analysis: Real World Applications}

\begin{document}

\begin{frontmatter}



\title{A new surrogate data method for nonstationary time series}


\author[dgl]{Diego L. Guarín López\corref{cor1}\fnref{fn1}}
\ead{dlguarin@gmail.com} \cortext[cor1]{Corresponding author.}
\fntext[fn1]{Diego L. Guarín López is supported by the Universidad
Tecnológica de Pereira and Colciencias, Contract No 5304 of 2010.}

\author[dgl]{Alvaro A. Orozco Gutierrez}
\ead{aaog@utp.edu.co}

\author[edt]{Edilson Delgado Trejos}
\ead{edilsondelgado@itm.edu.co}

\address[dgl]{Department of Electrical Engineering. Universidad tecnológica of Pereira. Pereira, Colombia. }
\address[edt]{Research center at the Instituto Tecnológico Metropolitano. Medellín, Colombia.}
\begin{abstract}
Hypothesis testing based on surrogate data has emerged as a popular
way to test the null hypothesis that a signal is a realization of a
linear stochastic process. Typically, this is done by generating
surrogates which are made to conform to autocorrelation (power
spectra) and amplitude distribution of the data (this is not
necessary if data are Gaussian). Recently, a new algorithm was
proposed, the null hypothesis addressed by this algorithm is that
data are a realization of a non stationary linear stochastic
process, surrogates generated by this algorithm preserve the
autocorrelation and local mean and variance of data. Unfortunately,
the assumption of Gaussian amplitude distribution is not always
valid. Here we propose a new algorithm; the hypothesis addressed by
our algorithm is that data are a realization of a nonlinear static
transformation of a non stationary linear stochastic process.
Surrogates generated by our algorithm preserve the autocorrelation,
amplitude distribution and local mean and variance of data. We
present some numerical examples where the previously proposed
surrogate data methods fail, but our algorithm is able to
discriminate between linear and nonlinear data, whether they are
stationary or not. Using our algorithm we also confirm the presence
of nonlinearity in the monthly global average temperature and in a
small segment of a signal from a Micro Electrode Recording.
\end{abstract}


\end{frontmatter}


\section{Introduction}
\label{Introduction} Surrogate data method, initially introduced by
\citet{theiler} is nowadays one of the most popular tests used in
nonlinear time series analysis to investigate the existence of
nonlinear dynamics underlying experimental data. The approach is to
formulate a null hypothesis for a specific process class and compare
the system output to this hypothesis. The surrogate data method can
be undertaken in two different ways: \emph{Typical realizations} are
Monte Carlo generated surrogates from a model that provides a good
fit to the data; \emph{constrained realizations} are surrogates
generated from the time series to conform to certain properties of
the data. The latter approach is preferable for hypothesis testing
due to the fact that it does not requiere a pivotal statistics
\cite{theiler2}. In order to test a null hypothesis at a level of
significance $\alpha$, one has to generate $1/\alpha -1$
($2/\alpha-1$) surrogates for a one side (two side) test. Then, one
simply evokes whatever statistic is of interest and compares the
value of this statistic computed from data to the distribution of
values elicited from the surrogates. If the statistic value of the
data deviates from that of the surrogates, then the null hypothesis
may be rejected.
Otherwise, it may not.\\
The classical methods for constrained realizations named (i) Random
shuffle (RS); (ii) Random phase (RP); and, (iii) Amplitude adjusted
Fourier transform (AAFT) surrogates \cite{theiler}, were developed
to test the null hypothesis that the data came from a (i) i.i.d
gaussian random process, (ii)  linear correlated stochastic process;
and (iii) nonlinear static transformation of a linear stochastic
process. Surrogates generated with the RS method preserves the
amplitude distribution (AD) of the original data, while the ones
generated with the RP algorithm preserve the autocorrelation (AC)
and surrogates generated with the AAFT algorithm preserve both the
AD and the AC of the original data (in general this is not true,
this is why an improved version
of the AAFT algorithm was presented, referred as iAAFT \cite{Schreiber1996}).\\
Recently, \citet{Richard2010} showed that surrogates generated with
the mentioned methods are stationarized versions of the original
data. This imply that while the statistical properties of the data
might be time dependent, the statistical properties of the
surrogates will not. Because of this, when data becomes from a
non-stationary process,  it is impossible to make a statistical
comparison of data with its surrogates. So, the classical surrogate
data methods are not applicable to non-stationary process. Due the
importance of this kind of process, many modifications of the
classical methods have been presented. The first one can be
attributed to \citet{Schreiber2000}, in this approach the surrogate
data preserves the AC and any other desired property of the original
time series. To generate a surrogate one starts by random shuffling
the data, then measuring (for example) the AC of the surrogates and
defining an error function as the square difference of data AC minus
the surrogate AC. One has to keep permuting pairs until the error
function is minimized. To generate surrogates for non stationary
time series, one has to ensure that the surrogates also preserve the
local mean and variance of the data. This procedure can be done
iteratively by means of any optimization algorithm, but there is no
guarantee that one will not be stuck in a local minimum (this issue
was overcome in \cite{Schreiber2000} by using the simulated
annealing optimization method). Unfortunately, this method requires
a lot of computational time, so it is of limited applicability.\\
Recently, \citet{tomomichi2} presented a modification of the RP
method which makes it suitable for non-stationary data, they called
its method Truncated Fourier Transform (TFT). Surrogates generated
with the TFT algorithm are constrained to preserve the AC and the
local mean and variance of data so, surrogates will be
non-stationary if original data are non-stationary. Through this
method it is possible to test the null hypothesis that the data came
from a non-stationary linear correlated stochastic process. Since
surrogates generated with this method do not preserve the AD of
data, further hypothesis (e.g, data are a realization of a nonlinear
statical transformation of a non-stationary linear correlated
stochastic process) can not be tested. The aim of this paper is to
present a new surrogate data method through which is possible to
obtain surrogates that are constrained to preserve the AC, AD and
local mean and variance of data, but are otherwise random.\\ This
document is organized in the following way; initially we briefly
introduce the RP, AAFT and the TFT methods, followed by an
introduction to our method, named Amplitude Adjusted Truncated
Fourier Transform (AATFT). Then we introduce a methodology to accept
or reject a null hypothesis and proceed to apply the methods to
several simulated and real time series, showing the utility of each
one. Finally we present some concluding remarks.
\section{Surrogate data methods} \label{surrogate data methods}
As mentioned, the surrogate data methods, originally introduced by
\citet{theiler}, has become a very popular method for hypothesis
testing. The original algorithms can be stated as follows:
\subsection{The existing algorithms}
\subsubsection{Random Phase surrogates (RP)}\label{RP} The
surrogate data is generated by the following procedure:
\begin{enumerate}
   \item Start with the original data $x[t]$, $t=1,\cdots,N$.
  \item Compute $z[n]$, the Fourier transform of $x[t]$.
  \item  Randomize the phases: $z'[n]=z[n]e^{\imath \phi
  [n]}$.\\  Where $\phi[1]=0$ and $\phi[n]\in\mathcal{N}(0,2\pi)$, $n=2,\ldots, N$.
  \item Symmetrize $z'[n]$ (to obtain a real inverse Fourier Transform):\\
    $z'[n-i+1]=\widehat{z}'[i+1]$, $i=1, \ldots, floor(n/2)$,\\
    if $N$ is even then $z'[n/2 +1]=abs(z'[n/2 +1])$.\\
$\widehat{z}[n]$ is the complex conjugate of $z[n]$.
  \item Obtain $x'[t]$, the inverse Fourier transform of $z'[n]$.
\end{enumerate}
$x'[t]$ is the surrogate data of $x[t]$.\\
The surrogates maintain the linear correlation of the data, but by
means of the phases randomization, any nonlinear structure is
destroyed.
\subsubsection{Amplitude Adjusted Fourier Transform surrogates (AAFT)}\label{AAFT}
The surrogate data is generated by the following procedure:
\begin{enumerate}
\item Start with the original data $x[t]$, $t=1,\ldots,N$.
  \item Sort the data $Sx[k]$, $k=1,\ldots,N$.
  \item Compute $z[n]$, the Fourier transform of $x[t]$.
  \item  Make a ranked time series $Rx[t]$ defined to satisfy
         $Sx[Rx[t]] =x[t]$.
  \item Create a random data set $g[t]$, $t =1, \ldots, N$.
  \item Sort  the  random  gaussian  number  $Sg[k]$,  $k=1, \ldots , N$.
  \item Define a new time series $y[t] =Sg[Rx[t]]$.
  \item Generate a surrogate time series $y'[t]$ from $y[t]$ using
        the RP algorithm.
  \item Make a ranked time series $Ry'[t]$ of $y'[t]$.
  \item The surrogate time series of $x[t]$ is given by $x'[t] =
        Sx[Ry'[t]]$.
\end{enumerate}
$x'[t]$ is the surrogate data of $x[t]$.\\
It is evident that this process achieves two aims: First, just as
with RP algorithm, the power spectra (and therefore linear
correlation) of the data is preserved in the surrogate; and second,
the re-ordering process means that the AD of the data and surrogate
is also identical (this is actually not true, as this algorithm does
not simultaneously preserve both rank distribution and power
spectra, which is why the iAAFT \cite{Schreiber1996} has to be used
in most practical situations).
\subsubsection{Truncated Fourier Transform Surrogates (TFT)}\label{TFTS}
The TFT algorithm introduced a way to deal with non stationarity;
this algorithm works by preserving the low frequency phases in the
Fourier domain, but randomizing the high frequency components.\\ The
surrogate data is generated by the following procedure:
\begin{enumerate}
  \item Start with the original data $x[t]$, $t=1,\cdots,N$.
  \item Compute $z[n]$, the Fourier transform of $x[t]$.
  \item  Randomize the phases: $z'[n]=z[n]e^{\imath \phi
  [n]}$. Where\\
  $\phi[n]\in\mathcal{N}(0,\pi)$ if $n > f_{c}$. \\
  $\phi[n] = 0$ if $n \leq f_{c}$.
  \item Symmetrize $z'[n]$ (as in the RP algorithm).
  \item Obtain $x'[t]$, the inverse Fourier transform of $z'[n]$.
\end{enumerate}
$x'[t]$ is the surrogate data of $x[t]$.\\ While all phases are not
randomized in this method, it is possible to discriminate between
linearity and non-linearity because the superposition principle is
valid only for linear data. i.e., when data are nonlinear, even if
the power spectrum is preserved completely, the inverse Fourier
transform data
using randomized phases will exhibit a different dynamical behavior.\\
The surrogate data generated by this method are influenced primarily
by the choice of frequency $f_{c}$. If $f_{c}$ is too high, the TFT
surrogates are almost identical to the original data. In this case,
even if there is nonlinearity in irregular fluctuations, one may
fail to detect it. Conversely, if $f_{c}$ is too low, the TFT
surrogates are almost the same as the linear surrogate and the
long-term trends are not preserved. In this case, even if there is
no nonlinearity in irregular fluctuations, one may wrongly judge
otherwise. The method for selecting the correct value of $f_{c}$ was
presented in \cite{tomomichi2}.
\subsection{A new algorithm}
\subsubsection{Amplitude Adjusted Truncated Fourier Transform
surrogates (AATFT)}\label{AATFT} Surrogates generated with the TFT
algorithm do not preserve AD of data (this is actually not true, if
$f_{c}$ is high enough the surrogates AD will eventually be like the
data AD, but this imply that surrogates are too similar to data). It
is tempting to think that this issue can be overcome by simply
applying a similar procedure to the AAFT (or the iAAFT) algorithm,
but the solution is not so simple. The idea of the TFT method is to
preserve the low frequency components of data in surrogates, this is
done by preserving some phases of frequency domain, and it is
possible to observe that thanks to the reordering procedure of the
AAFT method
the phases will no longer be preserved.\\
In order to preserve the AC, AD and local mean and variance of data
in surrogates we propose the following procedure.
\begin{enumerate}
  \item Start with the original data $x[t]$, $t=1,\ldots,N$.
  \item Sort the data $Sx[k]$, $k=1,\ldots,N$.
  \item Compute $z[n]$, the Fourier transform of $x[t]$.
  \item Generate a surrogate time series $x'[t]$ of $x[t]$
  using the TFT algorithm.
  \item Compute $z'[n]$, the Fourier transform of $x'[t]$.
  \item Change the magnitude of $z'[n]$:\\
         $z'[n]= \left(z'[n]/abs(z'[n])\right)abs(z[n])$.
  \item Obtain $x'[t]$, the inverse Fourier transform of $z'[n]$.
  \item Make a ranked time series $Rx'[t]$ of $x'[t]$.
  \item Modify $x'[t]$ so it has the same data as $Sx[k]$ but
  with the order given by $Rx'[t]$: $x'[Rx'[t]]=Sx[k]$.
\end{enumerate}
$x'[t]$ is the surrogate data of $x[t]$.\\
If one iteratively performs the steps 5 to 9 it is possible to
increase the fitness between AC of the data and the surrogates (the
iterative procedure will also reduce the preservation of local mean
and variance, but this can be solved increasing the value of
$f_{c}$). This iterative procedure will be referred to as iAATFT.
Note that surrogate time series $x^{\ '}[t]$ is just a shuffling
of original time series $x[t]$, so it has the same AD.\\

It is important to notice that any implementation of the discrete FT
assumes that the time series under consideration is periodic with a
finite period. When there is a large difference between the first
and last points (end-point mismatch), the FT will treat this as a
sudden discontinuity in the time series. As a result, this will
introduce significant spurious high-frequency power into the power
spectrum, which is a critical problem when the randomization is
centered only on the high-frequency portion.\\ To ameliorate this
artifact, \citet{tomomichi2} proposed to symmetrize the original
data before the application of the FT (i.e.
$\{x_1,x_2,\cdots,x_{n-1},x_n,x_n,x_{n-1},\cdots,x_2,x_1\}$). With
this procedure, there is no end-point mismatch in the data.
\section{Testing for nonlinearity} \label{testing for nonlinearity}
Next we describe our selection of discriminant statistics, and
propose a methodology to accept or reject a null hypothesis using
this statistic. Finally we study a method for selecting the correct
value of $f_{c}$.\\ It is important to clarify that we are not
interested in performing a deep analysis on the linearity or
nonlinearity of any specific time series, our aim is to preset and
study the behavior of the new surrogate data method called AATFT;
further applications will be presented in the future.
\subsection{Selection of the discriminant statistics} Dynamical measures are often used as
discriminating statistics. According to \cite{bookthree}, the
correlation dimension  is one of the most popular choices. To
estimate these, we first need to reconstruct the underlying
attractor. For this purpose, a time-delay embedding reconstruction
is  usually applied. But this method is not useful for data
exhibiting irregular fluctuations and long-term trends, because a
smaller time delay is necessary to treat irregular fluctuations and
a larger time delay is necessary to treat long-term trends. At the
moment, there is no optimal method for embedding
such data \cite{bookthree}.\\
Therefore, as discriminant statistics we chose the Average Mutual
Information (AMI). The AMI is a nonlinear version of the AC.  It can
answer the following question: On average, how much does one learn
about the future from the past?. For further information regarding
the AMI, the reader is referred to \cite{bookthree} and references
within.
\subsection{Rejection or acceptance of a null hypothesis}
Surrogate data methods are based on the Monte Carlo hypothesis
testing procedure, first one calculates the statistic for data and
surrogates, and then one compares the value of this statistic
computed from the data to the distribution of values elicited from
the surrogates. If there is sufficient difference the null
might be rejected, otherwise it will not. \\
The level of significance of the test is given by the number of
surrogates. For a one sided (two sided) test a level of significance
$\alpha$ is reached with $1/\alpha - 1$ ($2/\alpha - 1$) surrogates.\\
For robustness, the AMI must be calculated for lag of 1. This
calculates on average how much information we have about
$\{x_{t+1}\}$ knowing $\{x_{t}\}$. So, if $AMI(\tau=1)$ of the data
deviates from that of the surrogates (i.e. is greater or lower) then
the null hypothesis may be rejected.\\
In order to reject (or not) a null hypothesis we generate $N=99$
surrogates, which gaves us a level of significance $\alpha = 0.02$.
\subsection{Selection of the correct value of $f_c$} The selection
of the correct value of $f_{c}$ cannot be done a priori, because it
depends on the nature of the data and the length of the time series
\cite{tomomichi2}. Our aim is to preserve AC, AD and local mean and
variances of the original data in the surrogates. The conservation
of AD is assured by the reordering process of the AATFT algorithm.
To preserve AC and local mean and variance one has to start by
randomizing all the phases ($100\%$ of frequency domain), if AC,
local mean and variance of surrogates are not similar to the data
then it is necessary to randomize only a portion of the phases (i.e.
$99\%$ of the higher frequency domain), and keep decreasing the
value of $f_c$ by small steps until the surrogates preserve AC,
local mean and variance of the original data. It should be noted
that it is no possible to preserve the AC for all lags, but at least
for small lags the surrogates AC should be identical to the data AC.
\section{Results}
\subsection{Numerical examples} In order to prove the validity
of our surrogate data method, we compared results obtained by
applying the different algorithms to a fictional time series.
\subsubsection{A simple example} 
\begin{figure}[ht]
\centering
  \includegraphics[width=8cm]{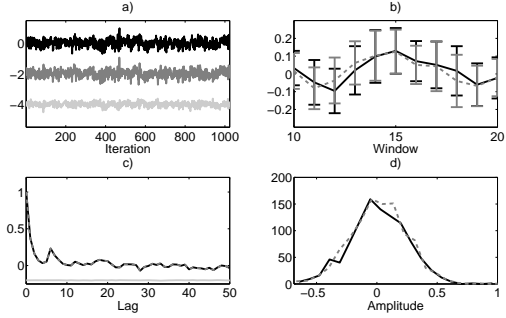}\\
  \caption{a) Original Data (black), surrogate data generated with the RP algorithm (dark gray) and difference between data and surrogate (gray).
   Values are displaced from one another by 2 for clarity. b) Local Mean and variance of  data (black) and  surrogates (dotted gray). c)
   AC of  data (black) and surrogates (dotted dark gray), and the difference between AC of data and surrogates (gray). The difference is
   displaced by 0.2 for clarity. d) AD of data (black) and of surrogates (dotted gray) }\label{fig1}
\end{figure}

\begin{figure}[ht]
\centering
  \includegraphics[width=8cm]{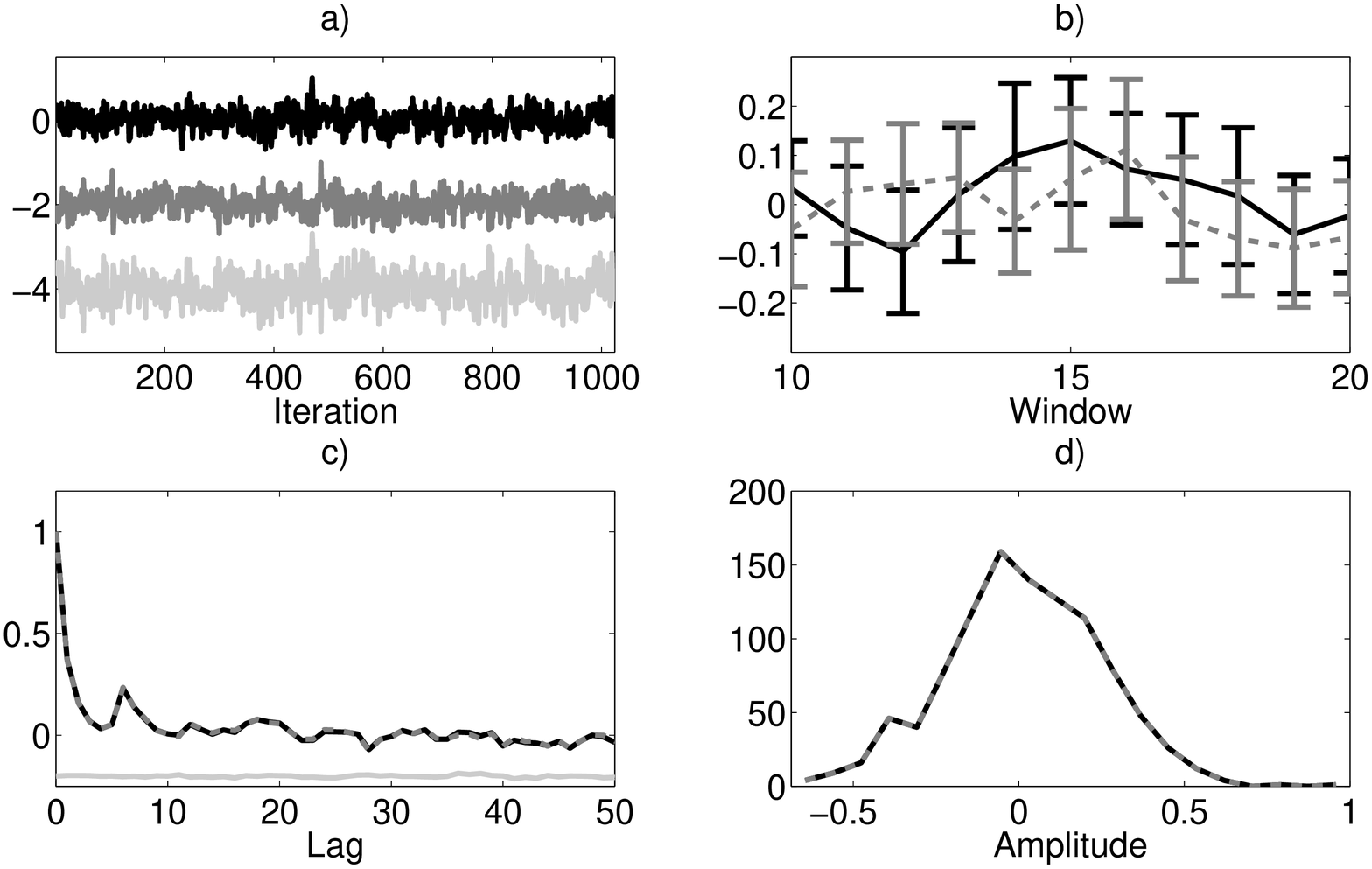}\\
  \caption{a) Original Data (black), surrogate data generated with the iAAFT algorithm (dark gray) and difference between data and surrogate (gray).
   Values are displaced from one another by 2 for clarity. b) Local Mean and variance of  data (black) and  surrogates (dotted gray). c)
   AC of  data (black) and surrogates (dotted dark gray), and the difference between AC of data and surrogates (gray). The difference is
   displaced by 0.2 for clarity. d) AD of data (black) and of surrogates (dotted gray) }\label{fig2}
\end{figure}

\begin{figure}[ht]
\centering
  \includegraphics[width=8cm]{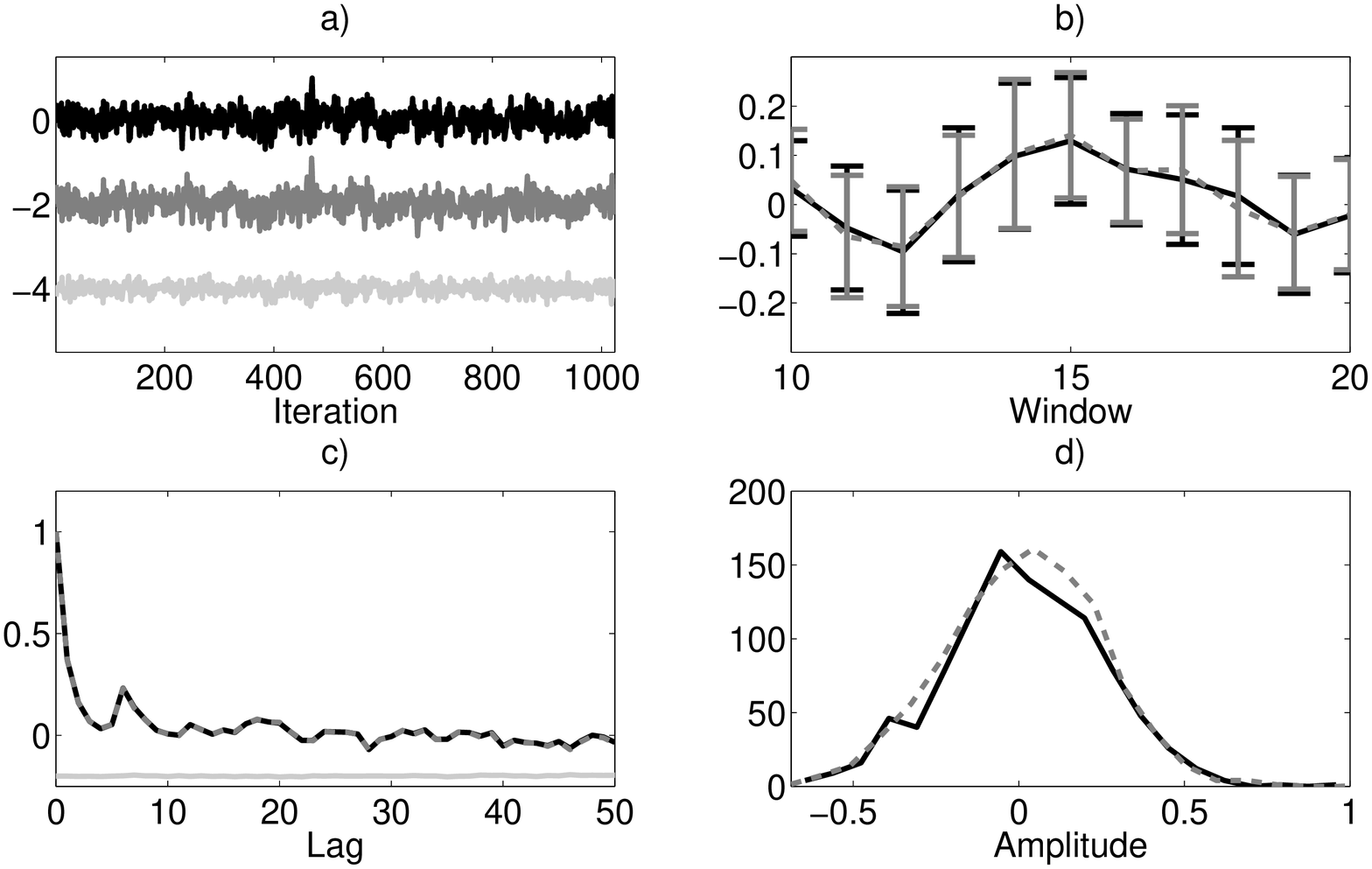}\\
  \caption{a) Original Data (black), surrogate data generated with the TFTS algorithm (dark gray) and difference between data and surrogate (gray).
   Values are displaced from one another by 2 for clarity. b) Local Mean and variance of  data (black) and  surrogates (dotted gray). c)
   AC of  data (black) and surrogates (dotted dark gray), and the difference between AC of data and surrogates (gray). The difference is
   displaced by 0.2 for clarity. d) AD of data (black) and of surrogates (dotted gray). In this case we randomized the higher $98\%$ of the frequency domain. }\label{fig3}
\end{figure}

\begin{figure}[ht]
\centering
  \includegraphics[width=8cm]{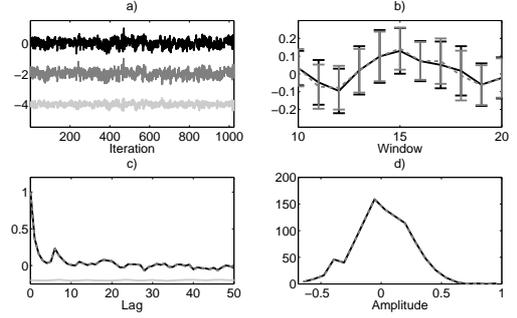}\\
  \caption{a) Original Data (black), surrogate data generated with the AATFT algorithm (dark gray) and difference between data and surrogate (gray).
   Values are displaced from one another by 2 for clarity. b) Local Mean and variance of  data (black) and  surrogates (dotted gray). c)
   AC of  data (black) and surrogates (dotted dark gray), and the difference between AC of data and surrogates (gray). The difference is
   displaced by 0.2 for clarity. d) AD of data (black) and of surrogates (dotted gray). In this case we randomized the higher $99\%$ of the frequency domain. }\label{fig4}
\end{figure}

First, we analyzed the data
generated by a linear AR model given by
\begin{equation}\label{eq1a}
x(t)=a_{1}x(t-1)+a_{6}x(t-6)+\eta.
\end{equation}
Where $a_{1}=0.3$, $a_{6}=0.2$ and $\eta \in \mathcal{N}(0,1)$. We
obtained 2048 values and discarded the first half. Our aim was to
prove what has been argued about each algorithm, in this case the
acceptance of the null hypothesis is granted. Figs. \ref{fig1} to
\ref{fig4} show that each algorithm achieves its goals, so using the
AATFT algorithm we can generate surrogate constrained to have the
same AC, AD and local mean and variance of the data.

\subsubsection{Failure of the iAAFT algorithm}
\begin{figure}[ht]
\centering
  \includegraphics[width=8cm]{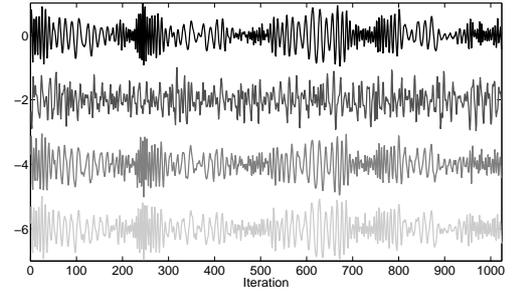}\\
  \caption{Data generated by a linear non stationary AR process (black), surrogate generated by the iAAFT algorithm (dark gray), the TFT algorithm (gray)
  and the iAATFT algorithm (light gray)  each displaced from the other by 2 units for clarity }\label{fig5}
\end{figure}

\begin{figure}[ht]
\centering
  \includegraphics[width=8cm]{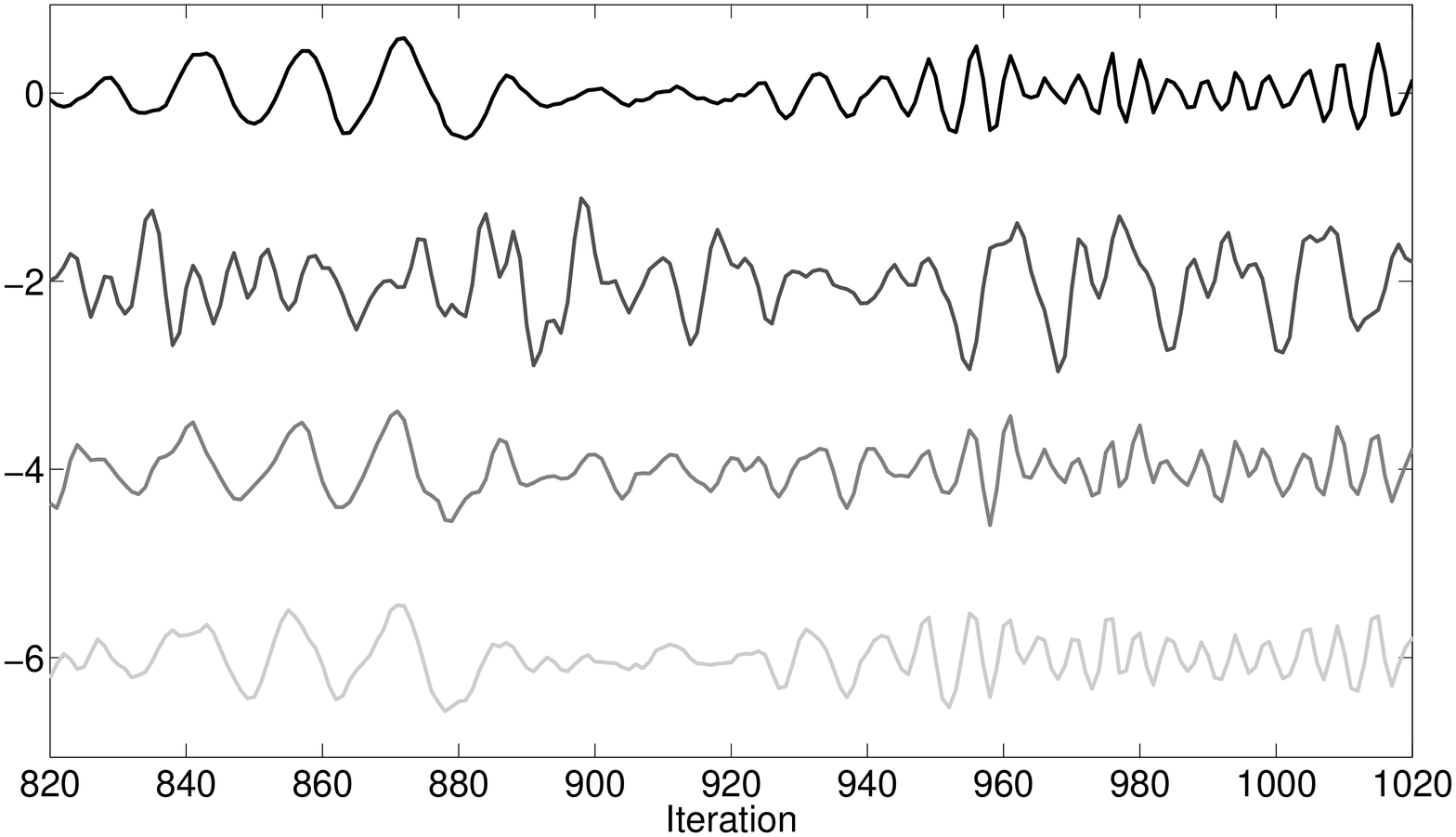}\\
  \caption{Data generated by a linear non stationary AR process (black), surrogate generated by the iAAFT algorithm (dark gray), the TFT algorithm (gray)
  and the iAATFT algorithm (light gray)  each displaced from the other by 2 units for clarity }\label{fig6}
\end{figure}

 To study the behavior of algorithms
in presence of non stationarity we followed \cite{Timmer1998}. First
we defined an AR process.
\begin{equation}\label{eq1}
 x(t)=a_{1}(t)x(t-1)+a_{2}(t)x(t-2)+a_{3}(t)+\eta.
\end{equation}
 Where,
\begin{equation}\label{eq2}
\begin{split}
    a_{1}& =2\cos{\left(2\pi /T\right)}\exp{\left(-1/\tau\right)},\\
    a_{2}&=-\exp{\left( -2 / \tau \right ) },\\
    a_{3}&=1.
\end{split}
\end{equation}
This process can be interpreted as a damped oscillator, with period
$T$ and relaxation time $\tau$. Period-based modulation is
introduced by subjecting the mean period of the AR process to a
sinusoidal fluctuation of the form
\begin{equation}\label{eq3}
    T(t)=T+M_{t}\sin{\left (t2\pi/T_{mod}\right)}.
\end{equation}
This modulation introduces a temporal dependency in $a_{1}$:
\begin{equation}\label{eq4}
     a_{1}(t)=2\cos{\left(2\pi /T(t)\right)}\exp{\left(-1/\tau\right)}.
\end{equation}
However, \ref{eq3} also introduces a temporal dependency in the
variance, which can be compensated by using
\begin{equation}\label{eq5}
\begin{split}
    a_{3}(t)^2=&\left(\frac{a_{3}^{2}}{1-a_{1}^2-a_{2}^2-2a_{1}^2a_{2}/(1-a_{2})}\right)\\
    &\times \left( 1-a_{1}(t)^2-a_{2}^2-\frac{2a_{1}(t)^2a_{2}}{1-a_{2}}
    \right).
\end{split}
\end{equation}
To generate data, we obtained 2048 values and discarded the first
half. Fig. \ref{fig5} shows the time series (the following
parameters were used: $T=50$,$\tau=10$, $T_{mod}=250$ and
$M_{T}=5.5$) and a surrogate generated with  each algorithm (we
excluded the RP algorithm).
Fig. \ref{fig6} shows an amplification of Fig. \ref{fig5}, it can be
noted that surrogates generated with TFT and iAATFT algorithms
preserve the low frequency behavior.
\begin{figure}[ht]
\centering
  \includegraphics[width=8cm]{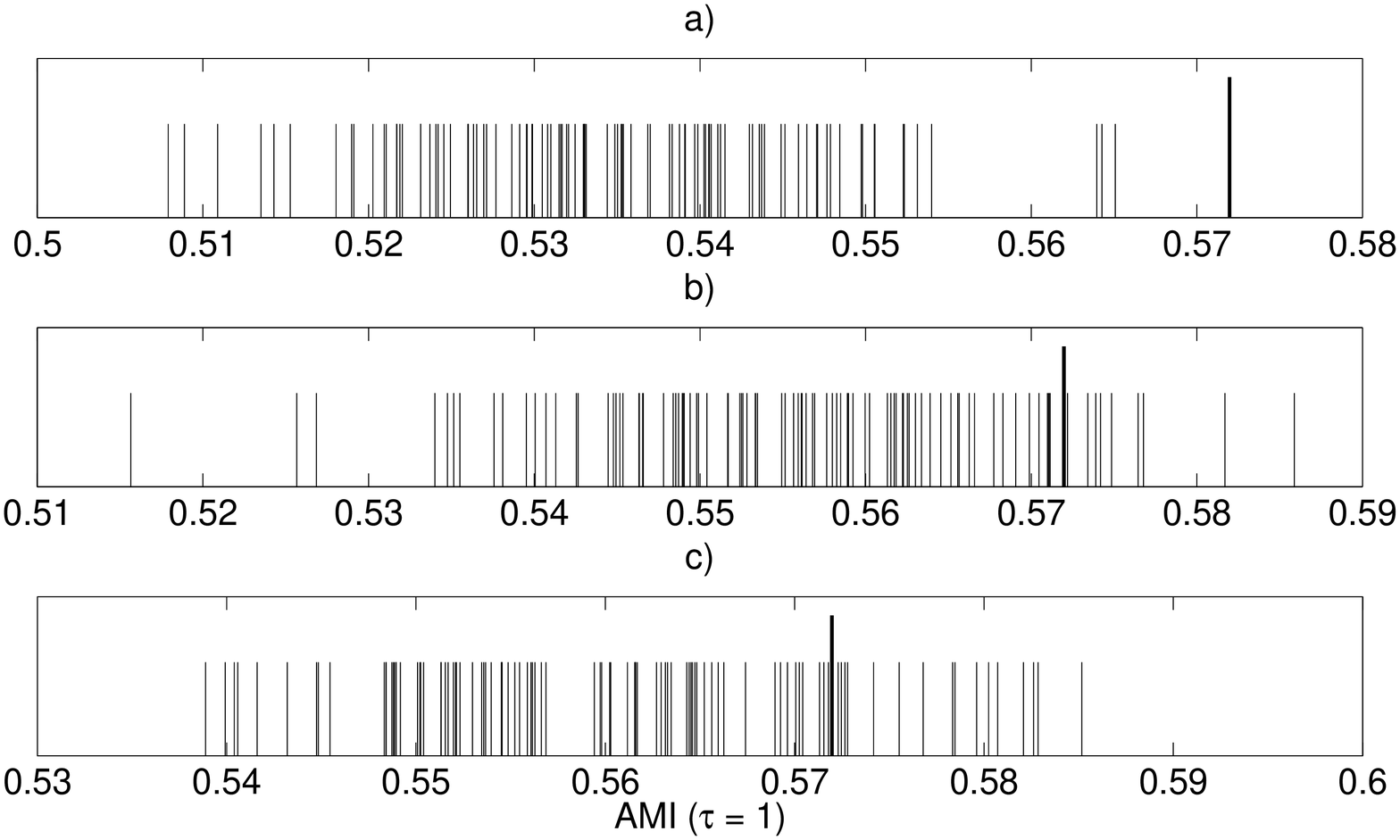}\\
  \caption{AMI ($\tau=1$) for data generated by a linear non stationary AR process (longer stem) and 99 surrogates generad with the a) iAAFT algorithm,
  b) TFT algorithm and c) iAATFT algorithm (10 iterations were performed). We randomized the higher $99\%$ of the frequency domain.}\label{fig7}
\end{figure}
Finally, Fig. \ref{fig7} shows the results of computing
$AMI(\tau=1)$ for data and $99$ surrogates generated with each
algorithm, it can be observed in Fig. \ref{fig7} a) that the null
hypothesis addressed by the iAAFT algorithm was rejected, but this
happens because the times series is non stationary not because it is
nonlinear. As expected, the hypothesis addressed by TST and AATFT
algorithms was not rejected (\citet{Timmer1998} proved that the
iAAFT algorithm is robust for some kinds of non-stationarity, but as
seen here this is not a general result).
\subsubsection{Failure of the TFT algorithm}
Next we generated surrogates for the following process
\begin{equation}\label{eq6}
    h(t)=g[x(t)]=x(t)^2.
\end{equation}
Where $x(t)$ is given by \ref{eq1}. In this case, the signal is
nonlinear, but the nonlinearity is given by the observation function
$g[ \ ]$ rather than by the dynamic of the process.
\begin{figure}[ht]
\centering
  \includegraphics[width=8cm]{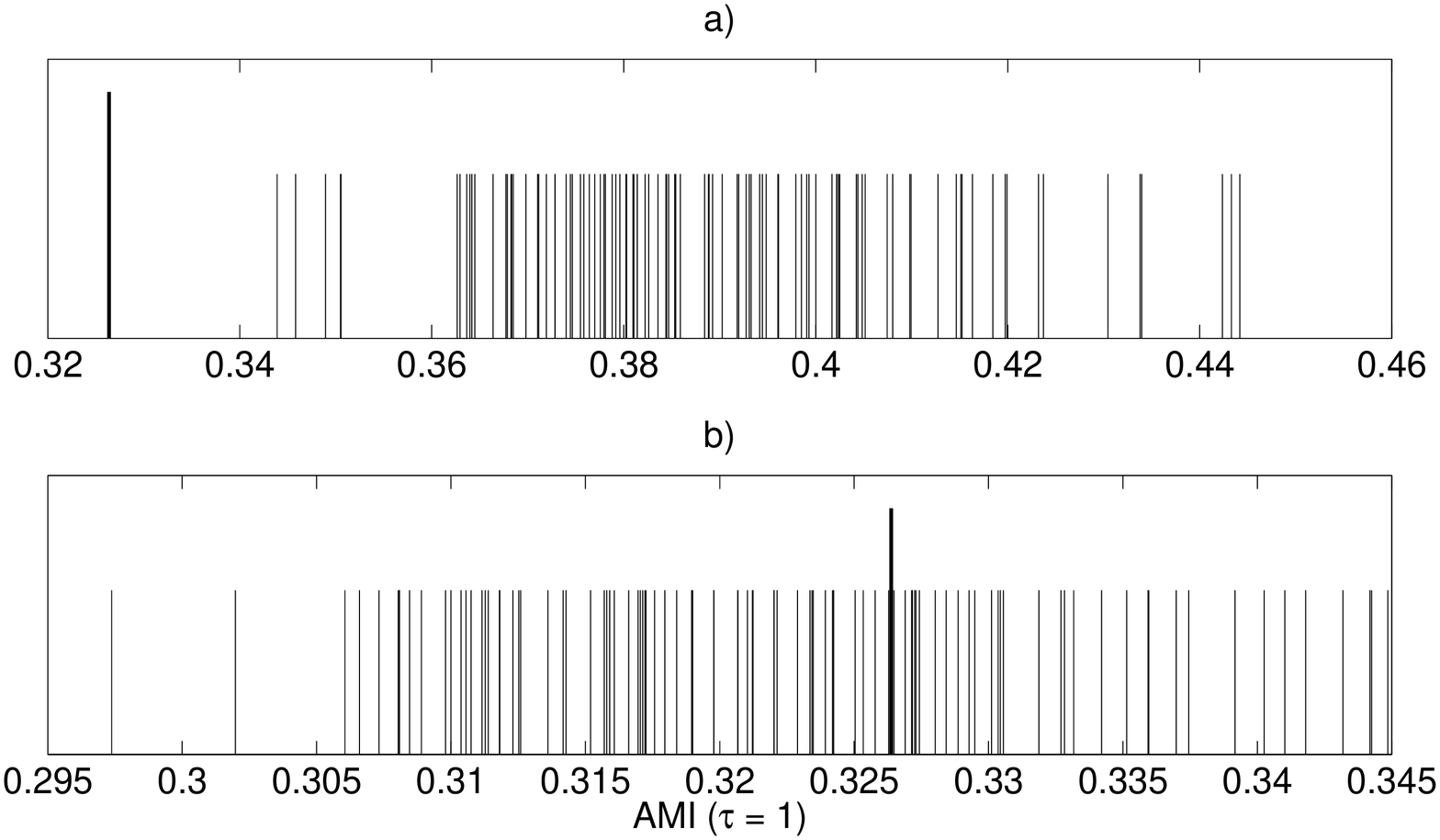}\\
  \caption{AMI ($\tau=1$) for data generated by a linear non stationary AR process observed through a nonlinear function (longer stem) and 99 surrogates generad with the
  a) TFT algorithm and b) iAATFT algorithm (10 iterations were performed). We randomized  the higher $98\%$ of the frequency domain.}\label{fig8}
\end{figure}
Fig. \ref{fig8} shows that the TFT algorithm detects nonlinearity,
but the iAAFT algorithm does not. This result was expected, because
the hypothesis addressed by the TFT algorithm does not involve a
static nonlinear transformation of the linear non stochastic
process, while this is exactly the hypothesis addressed by the AATFT
algorithm.
\subsubsection{Failure of the three methods} 
\begin{figure}[ht]
\centering
  \includegraphics[width=8cm]{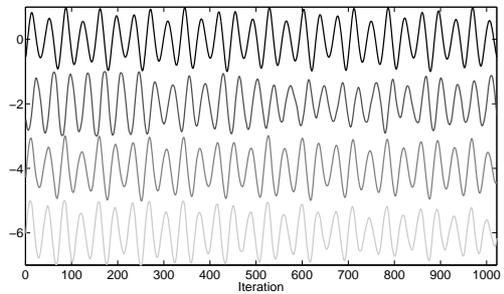}\\
  \caption{$x$ component of the Duffing equation (black), surrogate generated by the iAAFT algorithm (dark gray), the TFT algorithm (gray)
  and the iAATFT algorithm (light gray)  each displaced from the other by 2 units for clarity }\label{fig9}
\end{figure}

\begin{figure}[ht]
\centering
  \includegraphics[width=8cm]{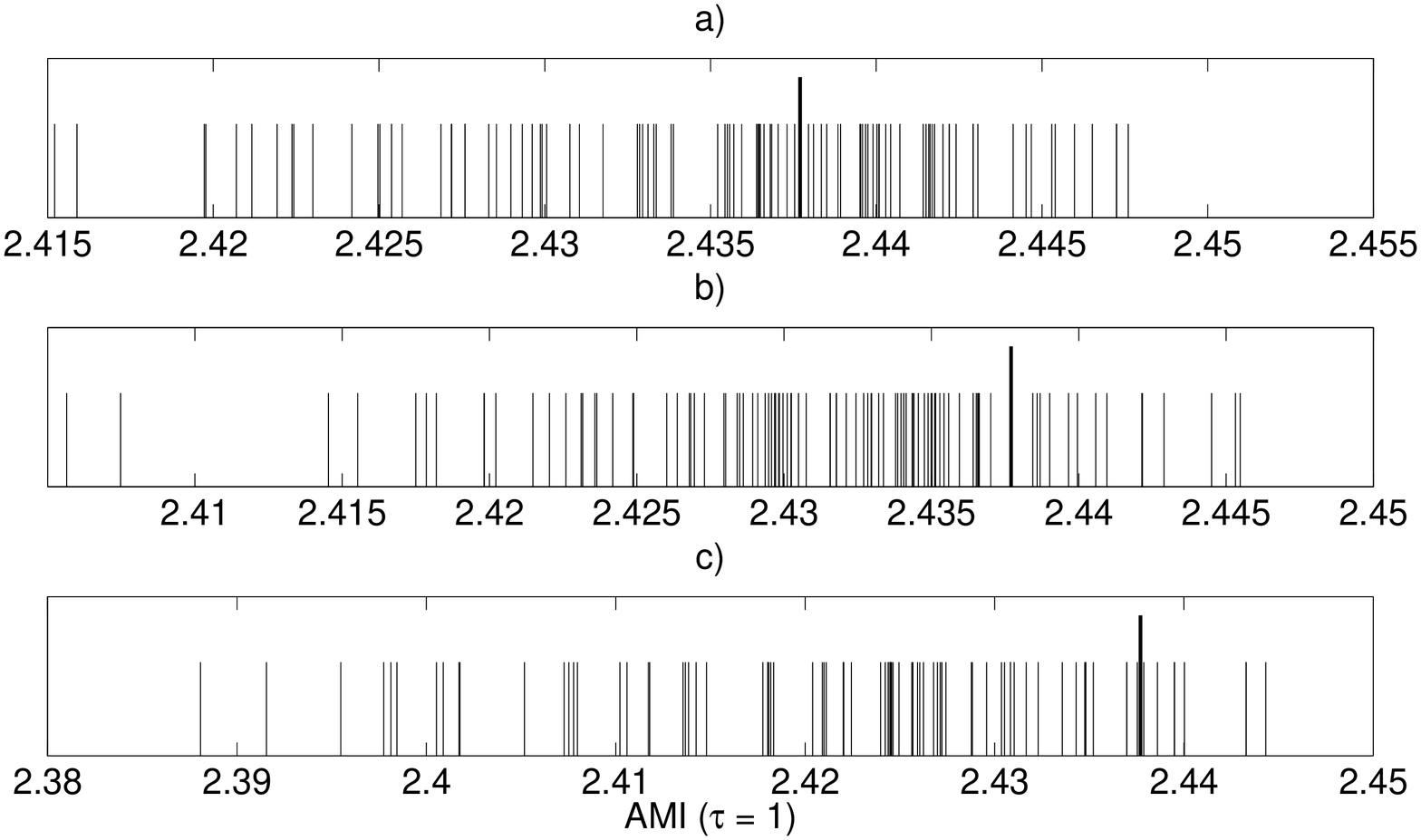}\\
  \caption{AMI ($\tau=1$) for $x$ component of the Duffing equation (longer stem) and 99 surrogates generad with the
  a) TFT algorithm and b) iAATFT algorithm (10 iterations were performed). We randomized the higher $95\%$ of the frequency domain.}\label{fig10}
\end{figure}

We now present a case
where neither of the hypotheses are rejected despite the fact that
the system that generated the signal is nonlinear. The signal was
generated by the Duffing system, given by
\begin{equation}\label{eq7}
\ddot{x}+\sigma \dot{x} +\omega_{0}^{2}x+\beta x = \gamma
\cos{\omega t}.
\end{equation}
In this case, $\sigma=0$, $\omega_{0}^{2}=\gamma=\omega=1$ and
$\beta=0.3$. The signal $x$ is obviously nonlinear, but this is a
case of weak nonlinearity \cite{Huang2005}.\\
Fig. \ref{fig9} shows 1024 points of the $x$ component of the
Duffing equation (integrated for $10.000$ steps with a unit of
$0.1$, discarded the first half and then selected a subsegment of
$1024$ points which minimized the end-point mismatch) and also shows
a surrogate generated with each algorithm. Surrogates generated with
each algorithm are very similar to the data, in this case we found
that randomizing the higher $95\%$ of the frequency domain, the AC,
local mean and variance of data were preserved in the surrogates
generated with the TFT and iAATFT methods.
Fig. \ref{fig10} shows that neither of the hypotheses can be
rejected using the AMI. The fact that the test fails to reject the
hypothesis could be a consequence of the selected discriminant
statistic or just because the nonlinearity is so weak that the test
simply fails to detect it.
\subsubsection{A chaotic system}
\begin{figure}[ht]
\centering
  \includegraphics[width=8cm]{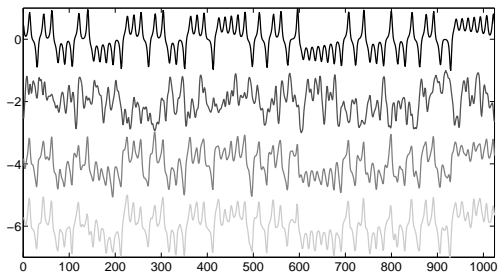}\\
  \caption{$x$ component of the Lorenz system (black), surrogate generated by the iAAFT algorithm (dark gray), the TFT algorithm (gray)
  and the iAATFT algorithm (light gray)  each displaced from the other by 2 units for clarity }\label{fig11}
\end{figure}

\begin{figure}[ht]
\centering
  \includegraphics[width=8cm]{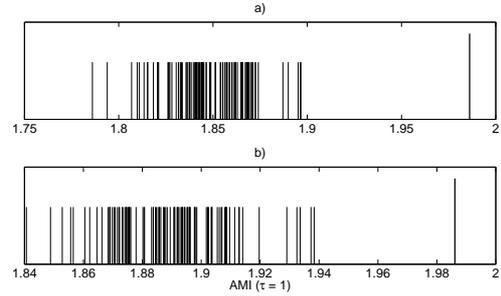}\\
  \caption{AMI ($\tau=1$) for $x$ component of the Lorenz system (longer stem) and 99 surrogates generad with the
  a) TFT algorithm and b) iAATFT algorithm (20 iterations were performed). We randomized the higher $97\%$ of the frequency domain.}\label{fig12}
\end{figure}

Subsequently, we used the iAAFT, TFT and the iAATFT to generate
surrogates for the Lorenz system \cite{bookthree}, which is given by
\begin{equation}\label{eq8}
\begin{split}
\dot{x}&=a(y-x)\\
\dot{y}&=x(b-z)-y\\
\dot{z}&=xy-cz
\end{split}
\end{equation}
The system exhibits a chaotic behavior with $a=10$, $b=28$ and
$c=8/3$.\\
Fig. \ref{fig11} shows 1024 points of the $x$ component of the
Lorenz system (integrated for $10.000$ steps with a unit of $0.1$,
discarded the first half and then selected a subsegment of $1024$
points which minimized the end-point mismatch), and also shows a
surrogate generated with each algorithm. It is easy to see that the
surrogate generated with the iAAFT algorithm is very different to
the data despite it preserve the AC and AD of the data (this
situation was also observed in Fig. \ref{fig5}), this imply that
data is either: nonlinear and stationary, linear and non stationary
or nonlinear and non stationary, but we cannot make a clear
distinction.
Fig. \ref{fig12} helps us clarify this issue, it is obvious that
data is nonlinear, because the linear and stationary and linear and
non stationary hypotheses were rejected.
\subsection{Application to real data} Based on the previous results,
we applied the TFT and the AATFT algorithms to two experimental
systems: (i) monthly global average temperature (MGAT) from January
1880 to February 2010 (1562 data points). This database is public,
available on the web and (ii) Micro electrode recording (MER) from
the substantia nigra pars reticulata (4096 data points) adquiared
during a Parkinson surgery held in Valencia (Spain). The equipment
used in the acquisition was the LEADPOINT TM of Medtronic.
\begin{figure}
\centering
  \includegraphics[width=8cm]{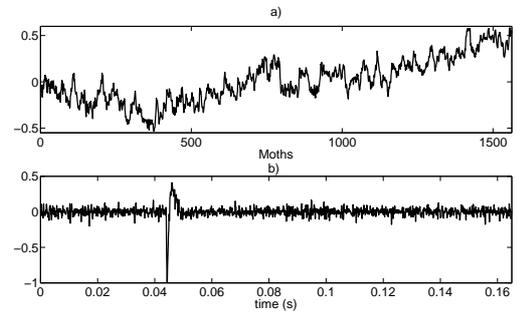}\\
  \caption{Real time series. a) Monthly global average
temperature from January 1880 to February 2010 and b) Micro
electrode recording (MER) from the substantia nigra pars
reticulata,}\label{fig13}
\end{figure}
\subsubsection{Monthly global average temperature (MGAT)} 
\begin{figure}[ht]
\centering
  \includegraphics[width=8cm]{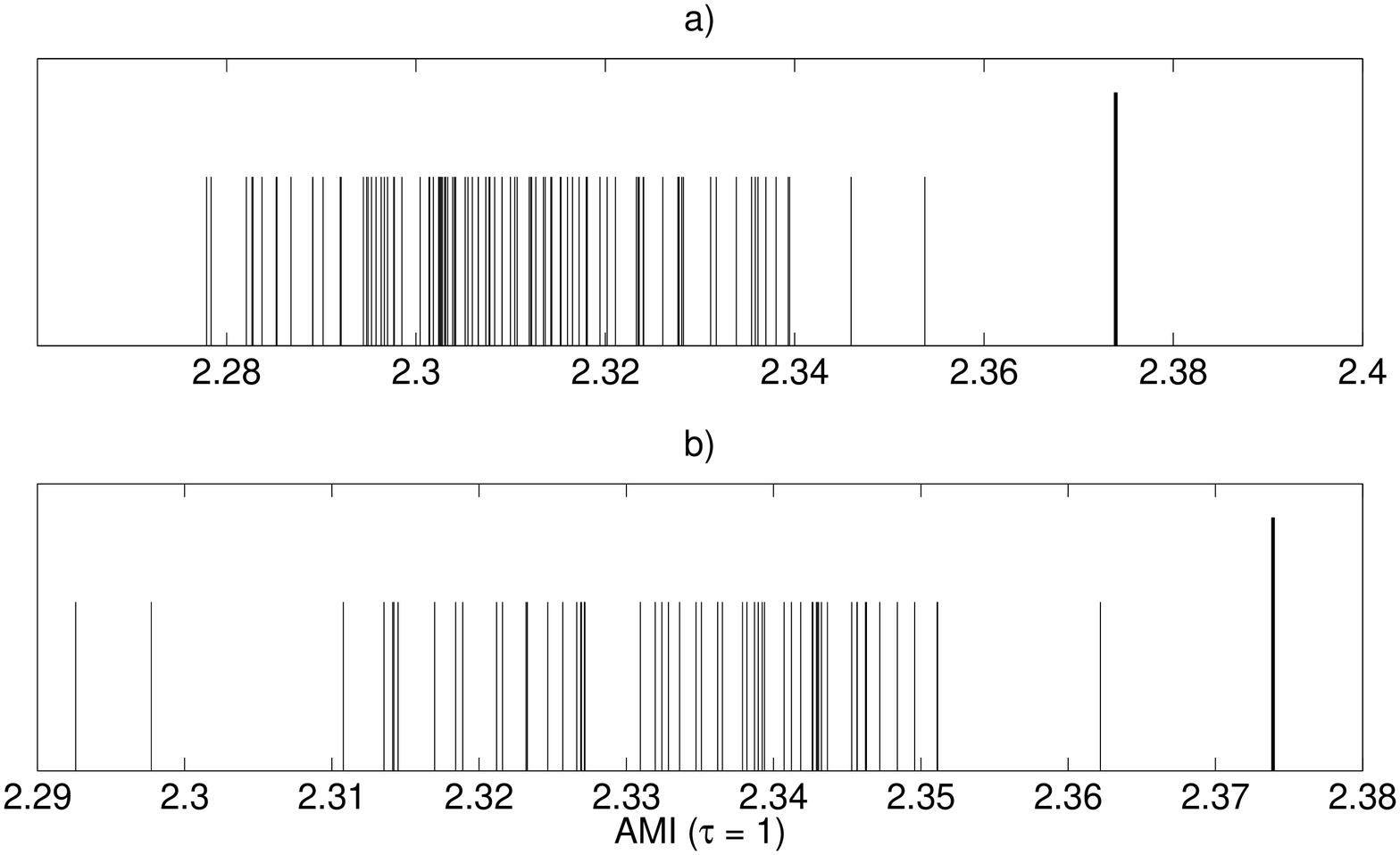}\\
  \caption{AMI ($\tau=1$) for the Monthly global average
temperature from January 1880 to February 2010 (longer stem) and 99
surrogates generad with the
  a) TFT algorithm and b) iAATFT algorithm (10 iterations were performed). We randomized the higher $98\%$ of the frequency domain.}\label{fig14}
\end{figure}

As shown in
Fig. \ref{fig13} a) the MGAT data is non stationary (it has a trend)
and has an end point mismatch, so the classical surrogate data
methods would not be able to detect nonlinearity. Prior to the
generation of surrogate data with the TFT and AATFT algorithms we
proceeded to symmetrize the data in order to eliminate the end point
mismatch. \\
Fig. \ref{fig14} shows that both hypotheses were rejected, so there
is a good chance that the data is nonlinear (there is always room
for error). This result verifies what was found by
\cite{tomomichi2}.
\subsubsection{Micro Electrode recordings (MER)}
\begin{figure}[ht]
\centering
  \includegraphics[width=8cm]{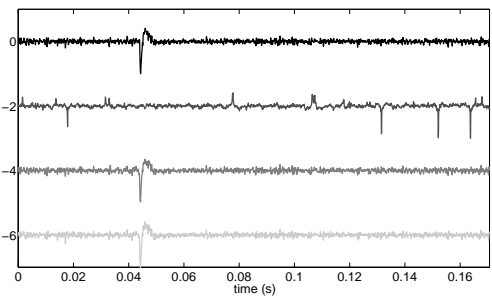}\\
  \caption{MER signal from the substantia nigra pars reticulata(black), surrogate generated by the iAAFT algorithm (dark gray), the TFT algorithm (gray)
  and the iAATFT algorithm (light gray)  each displaced from the other by 2 units for clarity }\label{fig15}
\end{figure}

\begin{figure}[ht]
\centering
  \includegraphics[width=8cm]{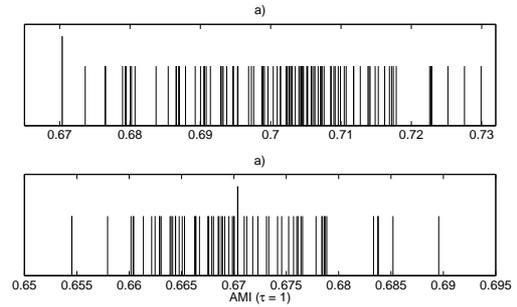}\\
  \caption{AMI ($\tau=1$) for MER signal from the substantia nigra pars reticulata (longer stem) and 99 surrogates generad with the
  a) TFT algorithm and b) iAATFT algorithm (10 iterations were performed). We randomized the higher $97.5\%$ of the frequency domain.}\label{fig16}
\end{figure}

We now turn our attention to physiological data. Fig. \ref{fig13} b)
shows the typical behavior of these kinds of signals, that is
synchronization; the spike is generated because of the
synchronization of a small cumulus of neurons sorrounding the
micro-electrode implanted in the brain, obviously this is a
difficult case and the standard surrogate data methods are useless.
However, the TFT and the proposed (AATFT) methods are able to mimic
the temporal behavior of data which implies  that the preservation
of local mean and variance is key to generating valid surrogates.
These results are shown in Fig. \ref{fig15}.
Finally, Fig. \ref{fig16} shows that the hypothesis addressed by the
TFT algorithm is rejected, while the hypothesis addressed by the
AAFT algorithm is not. This implies that data is nonlinear, but
nonlinearity is due to the observation function, further discussion
on this matter will be presented in the future.
\section{Conclusions}
A new surrogate data algorithm was presented. With this algorithm we
were able to generate surrogate data that are constrained to the
have the same autocorrelation (power spectra), amplitude
distribution and local mean and variance of data, but are otherwise
realizations of a non-stationary linear stochastic process. In this
way we expanded the range of uses for surrogate data methods, by
including non stationary and non Gaussian processes. Through
numerous examples we demonstrate that classical surrogate data
methods will fail to discrimine between linear and nonlinear systems
when the underlying process is non-stationary; we also shown that
the same problem occurs with the TFT surrogate method when the time
series generated by a  non-stationary process does not have a
Gaussian AD. Only the proposed AATFT algorithm is able to detect the
true nature of the data in this cases. \\
With these methods we were able to confirm the presence of
nonlinearity in the monthly global average temperature time series,
and we also prove that the studied MER signal is a realization of a
nonlinear statical transformation of a linear non-stationary
stochastic process, a result that will be studied further in future
works.
\section{ACKNOWLEDGMENTS}
We express our sincere gratitude to the Instrumentation and Control
Research Group of the Universidad Tecnológica of Pereira and to the
MIRP group in the Research Center of the Instituto Tecnológico
Metropolitano of Medellín - Colombia.




\bibliographystyle{model1-num-names}
\bibliography{bib}







\end{document}